\documentclass[runningheads]{llncs}
\usepackage[T1]{fontenc}
\usepackage{graphicx}
\usepackage{xcolor}
\usepackage[normalem]{ulem}
\usepackage{booktabs}

\begin{document}
\title{Enriching a Fashion Knowledge Graph from Product Textual Descriptions}
%
%
\author{João Barroca\and
Abhishek Shivkumar \and
Beatriz Quintino Ferreira \and Evgeny Sherkhonov \and João Faria}
\authorrunning{J. Barroca et al.}
%
\institute{Farfetch\\
\email{\{joao.barroca, abhishek.shivkumar, beatriz.quintino, evgeny.sherkhonov, joao.faria\}@farfetch.com}}
\maketitle              
\begin{abstract}

Knowledge Graphs offer a very useful and powerful structure for representing information, consequently, they have been adopted as the backbone for many applications in e-commerce scenarios.
In this paper, we describe an application of existing techniques for enriching the large-scale Fashion Knowledge Graph (FKG) that we build at Farfetch. In particular, we apply techniques for named entity recognition (NER) and entity linking (EL) in order to extract and link rich metadata from product textual descriptions to entities in the FKG. Having a complete and enriched FKG as an e-commerce backbone can have a highly valuable impact on downstream applications such as search and recommendations.
However, enriching a Knowledge Graph in the fashion domain has its own challenges. Data representation is different from a more generic KG, like Wikidata and Yago, as entities (e.g. product attributes) are too specific to the domain, and long textual descriptions are not readily available. Data itself is also scarce, as labelling datasets to train supervised models is a very laborious task. Even more, fashion products display a high variability and require an intricate ontology of attributes to link to.
We use a transfer learning based approach to train an NER module on a small amount of manually labeled data, followed by an EL module that links the previously identified named entities to the appropriate entities within the FKG. 
Experiments using a pre-trained model show that it is possible to achieve 89.75\% accuracy in NER even with a small manually labeled dataset. Moreover, the EL module, despite relying on simple rule-based or ML models (due to lack of training data), is able to link relevant attributes to products, thus automatically enriching the FKG.


 \and {Knowledge Graph} \and {E-commerce}
\end{abstract}
%
%
%

\section{Introduction}
\label{section:introduction}

Knowledge Graphs (KGs) have proven to be essential in various domains including e-commerce as they power many important business applications such as search and recommendation. It is no exception for Farfetch\footnote{Farfetch - www.farfetch.com}, a global e-commerce platform for luxury fashion, where we have recently embarked on the mission of building our own Fashion Knowledge Graph (FKG). Currently, the FKG contains entities like products, their attributes, such as color or material, categories, product lines, brands, aesthetics, etc., that in total amount to a few billion triples. 

\begin{figure}[t]
  \centering
    \includegraphics[width=0.9\textwidth]{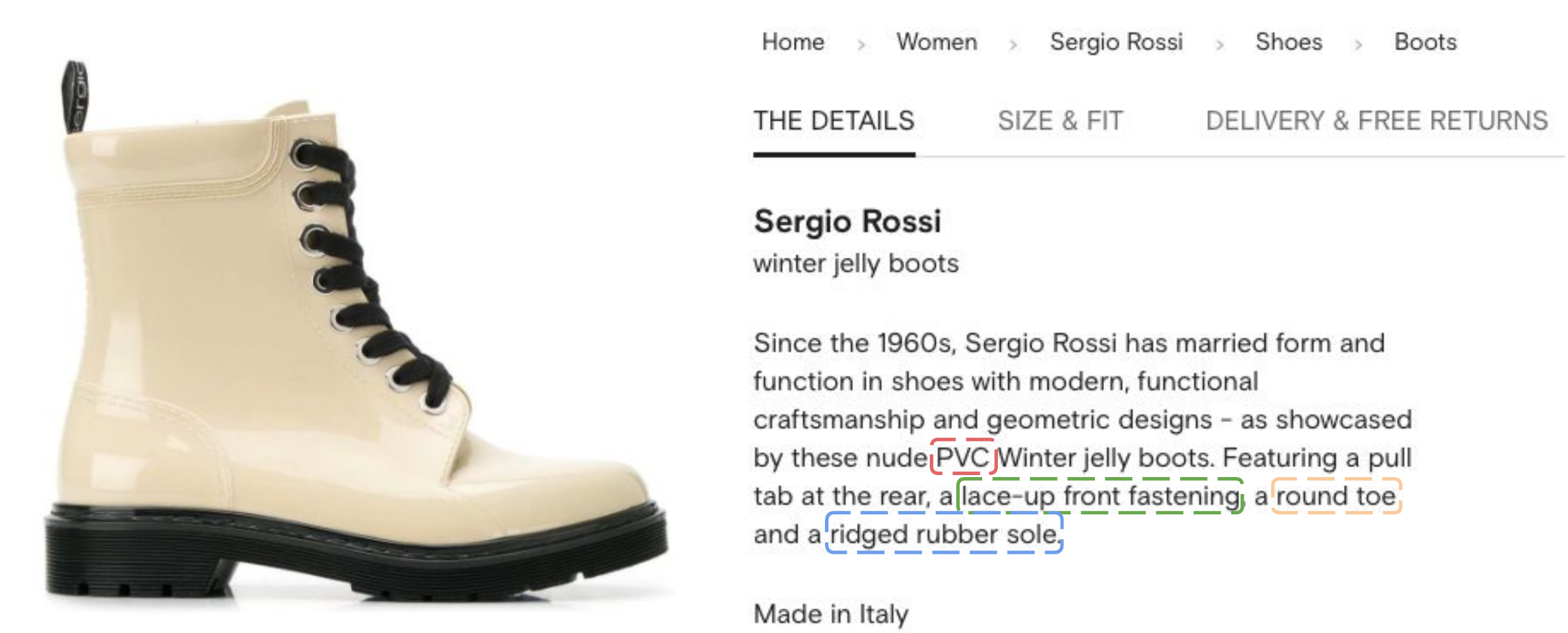}
      \caption{Example of a product that has no attributes specified but several attributes can be extracted from the description such as material, fastening type, toe shape, and sole type.}
    \label{example:product1}
\end{figure}

To keep the FKG up to date we have built scalable data pipelines that regularly update the graph with thousands of products and metadata so it always contains fresh and relevant data. The source data largely comes from structured data, e.g., internal data warehouses that are populated from web forms for product information that are filled in by the back office or sellers. As the latter process is manual and restricted to particular attribute types, it is often the case that products miss a lot of important information which leads to an incomplete KG, and thus to under-performing downstream applications.
Therefore, as part of our KG construction pipeline, we make use of Machine Learning models that extract product information from unstructured data such as images and text. For instance, for the product depicted in Figure~\ref{example:product1} the seller did not specify any attribute information, but these can be extracted either from the image, such as color information, or from the description, such as material, fastening type, toe type, and sole type. 	
To extract such information from product descriptions, as part of our FKG construction pipeline we have built Named Entity Recognition (NER) and Entity Linking (EL) modules. The NER module takes a product description as input, and tags sequences of strings, referred to as \emph{mentions}, with a possible attribute type. The EL module then links these tagged sequences to the existing attribute values that are stored in the FKG. For example, for the product in Figure~\ref{example:product1}, the NER module would tag the string "\texttt{PVC}" as \textsc{material}, then the EL module would link this mention to \texttt{:pvc}, an entity in the FKG of type \texttt{:Material}.

When building a KG for the e-commerce domain, we need to consider a few challenges. First, there is the sheer complexity of the domain, with thousands to millions of products types, organized into different taxonomy trees. In addition, there is a big variety across product types regarding the attributes. Not only the attributes are different when considering different types of products, but the attribute values vocabularies also vary across the different attributes. Together, these challenges result in a lack of labeled data for the full taxonomy of products in our catalog.

In this paper we demonstrate a simple approach for extracting mentions from product textual descriptions, linking them to existing attribute values stored in the FKG, and consequently enriching the products with new attribute metadata. As main contributions of this work, we:
\begin{itemize}
    \item demonstrate how we leveraged transfer learning to train an NER model, based on BERT~\cite{devlin2018bert} Transformer model, for the fashion domain with very limited labeled data;
    \item outline how simple rule-based models or traditional ML models using simple features can be used to build an EL system;
    \item report some enrichment statistics after applying the NER and EL components, end to end, to the textual descriptions of a set of products in our catalogue.
\end{itemize}



The rest of the paper is organized as follows. Section \ref{section:background} gives an overview of some of the relevant related works. In section \ref{section:method}, we describe the adopted methods, including a formalization of the problem followed by a detailed explanation of each method. Section \ref{section:experiments} lists all the experiments done, including details of the small in-house labeled dataset used to train the NER model. Lastly, section \ref{section:results} reports the experimental results and analysis, followed by the conclusions drawn in section \ref{section:conclusion}.


\section{Background and Related Work}
\label{section:background}

As previously introduced, our approach can be interpreted as combining two main steps: first a Named Entity Recognition (NER) task and, second, an Entity Linking task (EL). In the literature, NER can be either seen as a component of EL to detect the mentions in the text, or can be also handled independently from the EL system. At Farfetch, we had the manual resources available to annotate a small NER dataset and so we decided to have the NER as a separate independent component. In accordance, we will review the relevant state-of-the-art methods for NER and EL separately.

\subsection{Named Entity Recognition}
The Named Entity Recognition (NER) task has been widely used in the information extraction domain to identify entities, from unstructured text, and to classify them into pre-defined categories (like names of persons, organizations, or attributes).~\cite{NERsurvey2019} surveys long-established NER approaches, prior to transformers.

First approaches were commonly based on feature engineering and other supervised or semi-supervised learning algorithms, including Support Vector Machines (SVMs), Conditional Random Fields (CRFs) or Hidden Markov Models (HMMs)~\cite{NERsurvey2019}. 
With the ``deep learning'' revolution, a leap in benchmark performance was achieved by both word-level and/or character-level architectures based on applying a bidirectional-LSTM layer to word embeddings (namely \textit{word2vec} or \textit{glove}) and then obtaining a final prediction from a CRF layer~\cite{biLSTM-CRF2015,lample-etal-2016-neural}. LSTMs are very suitable for capturing the semantics and context of the entities as they are able to handle sparse context and have a sequential nature. On the other hand, LSTMs have been combined with CRFs to enforce tagging consistency and to extract cohesive and meaningful entities classifications. Attention mechanisms have also been applied (between the LSTM and CRF layers) in order to increase NER model's interpretability, i.e. to help explain the model’s decisions by highlighting the importance of key concepts relative to their neighborhood context, as in~\cite{opentag2018}. Particularly,~\cite{opentag2018,scalingopentag2019} are framed in the e-commerce domain and also use product textual information, such as title, descriptions, and bullet points. These works tackle the problem of Attribute Value Extraction (getting all attribute values for a set of products) and model it as a sequence tagging task, thus can be seen as a special case of NER. Marcelino et al., (2018) \cite{marcelino2018hierarchical} proposed a hierarchical deep learning natural language parser for fashion. Their approach not only recognized fashion-domain entities but also exposed syntactic and morphologic insights.

Transformers~\cite{Transformer2017} have indelibly changed the NER landscape. Current state-of-the-art NER methods are dominated by the Transformer architecture~\cite{Transformer2017} which handles sequential input data, like LSTMs, but solely relies on a self-attention mechanism. Such architecture brought significant performance and efficiency improvements and Transformers blocks have been extensively used for learning language representations applicable to many NLP downstream tasks. Popular Transformer models pre-trained on very large corpora, that have been fine-tuned to tackle the NER task, among others, include BERT~\cite{devlin2018bert}, that introduces a “masked language model” (MLM) pre-training objective, XLM-RoBERTa \cite{conneau-etal-2020-roberta}, which is also based on MLM but outperforms multi-lingual BERT on NER benchmarks. 
Employing a pre-trained (self-supervised) BERT-like model leverages the power of transfer learning in NLP, helping to solve the NER task for specific domains where data is scarce, as we only need to fine-tune it on the specific domain data.

\subsection{Entity Linking}

Entity Linking is the process of linking entity mentions appearing in unstructured text with their corresponding entities in a knowledge base (KB). Although Entity Linking can be applied to any kind of KB, in this review we will focus on methods that leverage the structured knowledge that we can find in knowledge graphs. 

Apart from NER, an entity linking system has other three main components: \textit{candidate generation}, which is responsible for retrieving candidate entities for each mention from a KB; \textit{candidate ranking/disambiguation}, which is responsible for scoring and/or ranking the candidate entities and choosing the one (or ones) to link; and, finally, \textit{unlinkable entities prediction} (also called NIL clustering) that deals with the mentions that were not linked to any entity.

Before focusing on specific relevant methods found in the literature for each of these components, and for the sake of completeness, we point the reader towards~\cite{NELDLsurvey1,NELDLsurvey2,NELKGsurvey2020}, where the first two surveys comprehensively review EL methods based on Deep Learning, while the latter presents an overview of recent advances in tasks for lifting Natural Language texts to KGs.

Approaches to \textit{candidate entity generation} have been primarily based on string matching techniques, both for named dictionary based techniques and for surface form expansion~\cite{ShenSurvey2015}.
In such techniques, string matching is performed between the mention string (or the surface form of the entity mention) and the name of the entity existing in the KB, usually using a named dictionary. Moreover, some entity linking systems try to leverage the whole Web information to identify candidate entities via Web search engines~\cite{ShenSurvey2015}.

The majority of proposed methods in the literature have either focused only on the \textit{candidate ranking/disambiguation} component or, more recently and with the advent of deep neural networks, are \textit{end-to-end joint approaches}. The latter approaches deal directly with an input text and aim to extract all candidate mentions and link them to their corresponding entities in a KB.

Within \textit{disambiguation} only approaches, Hoffart at al.~\cite{hoffart} unifies prior approaches into a comprehensive framework that combines three measures: popularity prior, similarity, and coherence. In addition, it also introduces new measures for defining mention-entity similarity, and a new algorithm for computing dense sub-graphs in a mention-entity graph, which produces high quality mention-entity mappings. However, the main contribution of~\cite{hoffart} is one of the best known benchmarks in the entity disambiguation task: the AIDA CoNLL-YAGO - a new dataset based on CoNLL 2003, in which they manually annotated all proper nouns with corresponding entities in YAGO2.

Recently, transformers have been widespreadly used in disambiguation tasks, leading to performance leaps on multiple benchmarks. 
Specifically, Yamada et al.~\cite{yamada} introduce a confidence-order model that achieves top results in several benchmarks, by making use of the recent pre-trained BERT model. In this work, the masked language model task is tackled as a masked entity prediction task.
Considering the context provided by KGs, Mulang et al.~\cite{mulang} demonstrate that pre-trained transformer-based models, although powerful, are limited to capturing context available purely on the texts concerning the original training corpus. They observe that adding an extra, task-specific, KG context improves the performance of Candidate Entity Disambiguation methods, leading to a new best performance for AIDA-CoNLL dataset.

As for \textit{end-to-end approaches}, relevant NN-based end-to-end linking models include: Ment-norm~\cite{letitov2018}, Kolitsas et al.~\cite{kolitsas2018}, NCEL~\cite{ycao2018}, and Martins et al.~\cite{martins2019}.
Ment-norm~\cite{letitov2018} achieved the best end-to-end results followed by Kolitsas et al.~\cite{kolitsas2018}.
Kolitsas et al.~\cite{kolitsas2018} use a bi-LSTM based model for mention detection and computes the similarity between the entity mention embedding and a set of predefined entity candidates. The authors demonstrate that engineered features are almost unnecessary when using end-to-end approaches. Their model reaches SOTA results on the AIDA/CoNLL dataset (comparing with the other end-to-end methods) and, when combined with Stanford NER, it generalizes well to other datasets with different characteristics.
Ment-norm~\cite{letitov2018} is an end-to-end system for NEL that considers relations as latent. Representation learning was used to learn relation embeddings, eliminating the need for extensive feature engineering.
Martins et al. (2019)~\cite{martins2019} also explore joint learning of Named Entity Recognition (NER) and EL showing that the two tasks benefit from joint training.
More recently, Cao et al.~\cite{cao} propose GENRE, a novel paradigm to address entity retrieval: it generates entity names autoregressively (leveraging pre-trained autoregressive language model BART and tackling EL as sequence-to-sequence problem). m-GENRE \cite{cao2} extends GENRE~\cite{cao} to multi-language. 
In~\cite{cao3}, Cao et al., make an improvement over the previous study (GENRE\cite{cao}) that boosts performance as it parallelizes autoregressive linking across all potential mentions and relies on a shallow and efficient decoder. Moreover, they add a new discriminative component that optimizes generators ranking.
Finally, Ravi et al.~\cite{ravi} propose CHOLAN, a modular approach to target end-to-end entity linking. CHOLAN consists of a pipeline of two transformer-based models integrated sequentially to accomplish the EL task. The first transformer model performs Named Entity Recognition in a given text and the second performs Named Entity Disambiguation (using the Wikipedia pages as entities external context).

Comparing disambiguation-only with end-to-end methods, taking into account benchmark results reported in the previously mentioned works, we observe that disambiguation-only methods usually outperform end-to-end methods in the entity disambiguation task. However, the recent GENRE~\cite{cao}, has competitive results for entity disambiguation and achieved SOTA in the entity linking task.
Nevertheless, end-to-end models are usually based on deep networks which need a huge amount of training data to generate robust models that perform well. In fact,~\cite{kolitsas2018} has shown that the accuracy of end-to-end models drops dramatically when trained on or applied to small datasets. This becomes a serious limitation especially when we're performing EL in specific (small) domains, as is the case of in work.

\section{Method}
\label{section:method}

Recently at Farfetch we embarked on the journey of building our own Fashion Knowledge Graph (FKG). The purpose of the FKG is to model and structure our knowledge about fashion and unify information about the main entities such as products, brands, aesthetics, etc. Such unified information can be effectively used in various applications such as search and recommendations.

The FKG is modelled as a set of RDF\footnote{\url{https://www.w3.org/TR/PR-rdf-syntax/}} triples, i.e., tuples of the form $(s, p, o)$, where $s$ and $p$, called the \textit{subject} and the \textit{property}, are an IRI, and $o$, called the \textit{object}, is an IRI, blank node, or a literal.
In the FKG, entities can be instances of different classes, such as \texttt{:Product} or \texttt{:Brand}, and be connected via properties such as \texttt{:hasBrand} and \texttt{:hasAesthetic}. Since in this paper we are interested in enriching the FKG with the product attributes extracted from the textual descriptions, we therefore enrich only a single property \texttt{:hasAttribute} that connects products to its attributes in the FKG. Rest of the data for FKG, on the other hand, comes from other internal or external data sources and enrichers.

In what follows, we outline the methods involved in extracting the attribute mentions from the textual product descriptions and linking them to the attribute in the FKG. Attribute \textit{mentions} are the sequence of strings extracted from the textual product descriptions that describe a possible attribute value. First, we define and formalize our task. Next, we detail our method for detecting attribute mentions in unstructured text. Finally, we explain how these mentions can be linked to the attribute values presented in the FKG.

\subsection{Task Definition}
\label{subsection:task_definition} 


Given a textual product description, our objective is to extract relevant product attributes. The \textit{relevant} attributes are the attributes presented in our Fashion Knowledge Graph.

We formulate this task as a two step process. The first step is a Named Entity Recognition (NER) component, which detects attribute mentions in the unstructured text. In addition, NER also detects the type of attribute associated with each mention. 
Formally, consider a product $P$ and its textual description $T^P=(t_1^P, t_2^P, ..., t_n^P)$ where $t_i^P$ refers to the $i^{th}$ token of the product description. The \emph{named entity recognition} task requires to detect a set $M^P = \{m_1^P, m_2^P, ..., m_k^P\}$, $k \leq n$, such that every $m_j^P$ is a pair of mention and mention type. In the following we often use the notions of mentions and mention pairs interchangeably.
For the example in Figure \ref{example:product1}, the NER detects the following attribute mention and mention type pairs: ("\texttt{PVC}", \textsc{material}), ("\texttt{lace-up front fastening}", \textsc{closure type}), ("\texttt{round toe}", \textsc{toe shape}), and ("\texttt{ridged rubber sole}", \textsc{sole type}).

The second step is an Entity Linking (EL) component, in which the detected attribute mention pairs are linked to a set of entities from the FKG. 
Formally, consider the previous detected attribute mentions $M^P$. In addition, we are given a set of attribute entity and type pairs $E = \{e_1, e_2, ..., e_l\}$ extracted from the FKG. The \emph{entity linking} task is the task of linking each mention $m_j^P\in M^P$ to a unique entity $e_i^P\in E \cup \{\texttt{NIL}\}$,  where \texttt{NIL} is a special null symbol to denote that no entity is linked.
As an example, consider the following set of entity and entity type pairs, corresponding to specific product attributes, retrieved from the FKG: \{(\texttt{:pvc}, \texttt{:Material}), (\texttt{:cotton}, \texttt{:Material}), (\texttt{:round}, \texttt{:Toe Shape})\}. The detected
mention pairs from the previous example would be linked with to following entities: 
\begin{itemize}
    \item ("\texttt{PVC}", \textsc{material}) $\rightarrow$ (\texttt{:pvc}, \texttt{:Material})
    \item ("\texttt{lace-up front fastening}", \textsc{closure type}) $\rightarrow$ \texttt{NIL}
    \item ("\texttt{round toe}", \textsc{toe shape}) $\rightarrow$ (\texttt{:round}, \texttt{:Toe Shape})
    \item ("\texttt{ridged rubber sole}", \textsc{sole type}) $\rightarrow$ \texttt{NIL}
\end{itemize}

\subsection{Named Entity Recognition}
\begin{figure}[tb]
    \centering
    \includegraphics[width=1.0\textwidth]{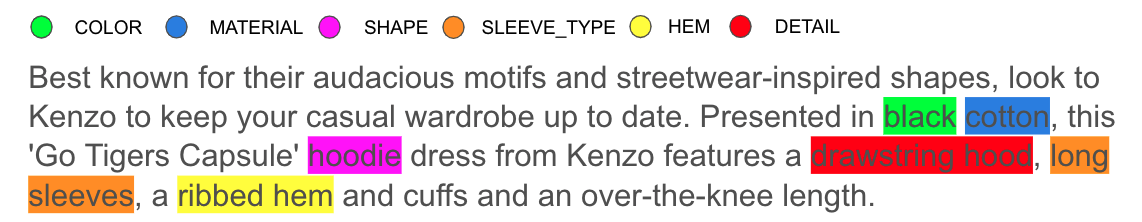}
    \caption{A sample product description showing the annotated entities.}
    \label{example:annotation_1}
\end{figure}

Due to the lack of labeled NER data within the Fashion domain, we make use of a pre-trained Transformer language model that has already been trained on a large amount of unstructured text. We take the weights from the pre-trained BERT-base-cased model~\cite{devlin2018bert} and fine-tune it on top of a small set of manually labeled NER dataset. We used an in-house built user-interface tool to manually label 900 descriptions of products from \textit{dresses} category across 11 classes.  We used the standard BIO \footnote{https://en.wikipedia.org/wiki/Inside-outside-beginning\_(tagging)} (short for beginning, inside, outside) notation for labelling the named entities, a common tagging format for tagging tokens in a chunking task in computational linguistics. The B-tag is used to indicate that the token is the beginning of a named entity phrase, while the I-tag is used for tokens inside the phrase, and the O-tag is used for tokens that do not belong to any named entity chunk. A sample product description with labeled named entities is shown in Figure~\ref{example:annotation_1}.  

\subsection{Entity Linking}
\label{subsection:entity_linking}

\begin{figure}[tb]
    \centering
    \includegraphics[width=1.0\textwidth]{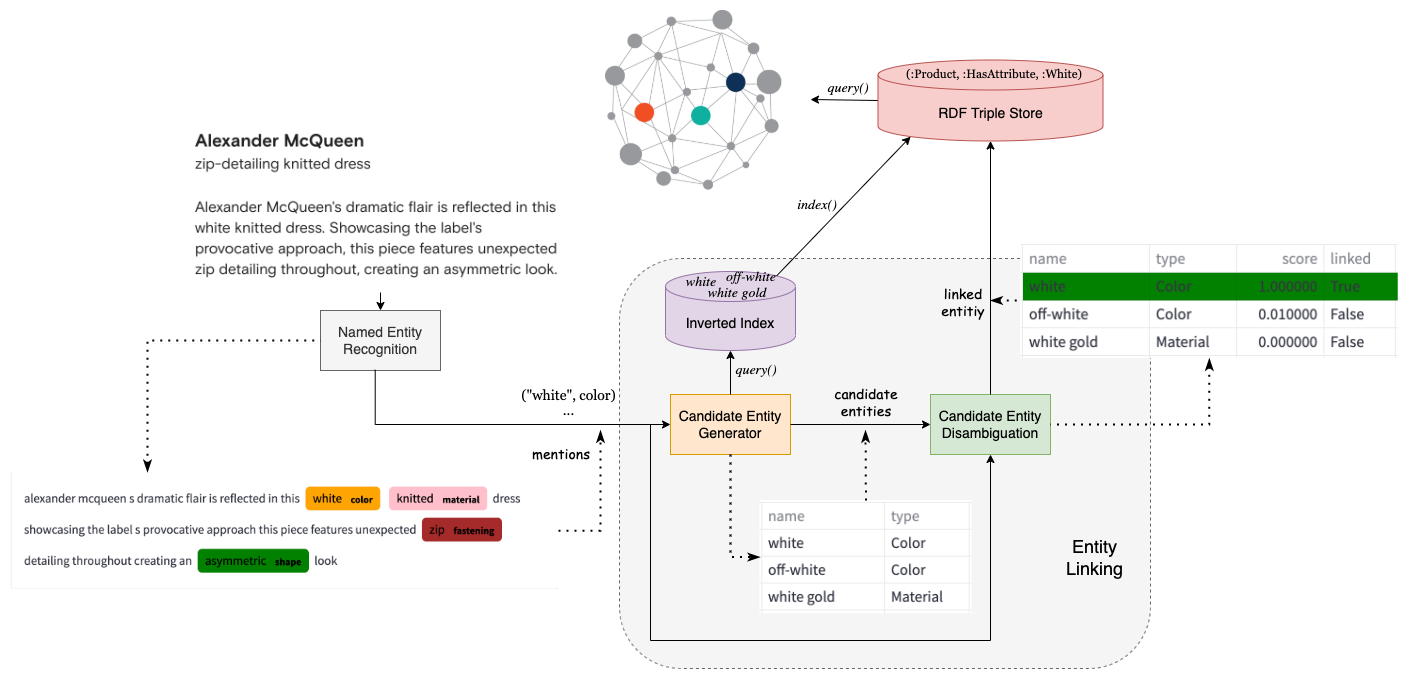}
    \caption{Full end to end system. Given a product textual description, the NER component detects attribute mentions and their corresponding types in the text. Next, for each attribute mention, the \textit{candidate entity generator} retrieves a set of candidate attribute entities. Lastly, the \textit{candidate entity disambiguation} predicts a score for each candidate, and decides which candidates should be linked or not. In the example described in the diagram, one of the attribute mentions detected is \texttt{"white"}, which is tagged as \texttt{color}. Given this mention, the following candidate entities are retrieved: (\texttt{:white}, \texttt{:Color}), (\texttt{:off-white}, \texttt{:Color}), and (\texttt{:white gold}, \texttt{:Material}). Since the entity (\texttt{:white}, \texttt{:Color}) is the one with higher score, and because the score is larger than a defined threshold, the entity is linked.}
    \label{fig:end2end_architecture}
\end{figure}

As explained in Section \ref{section:background}, traditional entity linking (EL) systems are composed of three main components: \textit{candidate entity generation}, \textit{candidate entity disambiguation} and \textit{NIL clustering}. In our solution, we only focus on the first 2 components. Figure \ref{fig:end2end_architecture} describes the full end-to-end system.

\subsubsection{Candidate Entity Generation} can be reformulated as an Information Retrieval (IR) problem: \textit{given a query, i.e. an attribute mention, we want to retrieve a set of results relevant to that query, i.e. a set of candidate entities}. Of the variety of IR techniques, the most widely used are based on the BM25~\cite{bm25} algorithm. BM25, where BM is an abbreviation of best matching, is a non-linear retrieval function that combines three key document attributes: term frequency, document frequency, and document length. There are a lot of available search engines that allow us to search for content based on inverted indexes, using the BM25 algorithm as a ranking function.

The attribute values in the FKG are represented as entities with labels as string literals. In order to be able to search over the attribute values in the FKG, we first create an inverted index for all their labels, and then use that index to perform full-text search using the BM25 function to rank the results.

\subsubsection{Candidate Entity Disambiguation} is formulated as a binary classification problem. Given a pair of an attribute mention and a candidate attribute entity, we predict if the candidate should be linked or not. We highlight, at this point, that we lack the training data needed to train more complex solutions based on Neural Networks. Therefore, we test two types of models: \textit{rule-based models} and \textit{traditional ML models}.

The rule-based models are based on domain-specific heuristics. First, we build a consistent text-preprocessing pipeline that enforces lower-casing, text normalization (including accents normalization and special characters removal), and stemming. Next, we create a type mapper which only allows certain types of entities to be linked with certain types of mentions. Finally, we build two different rule-based models. One of them performs an \texttt{exact-match} between the mention and the candidate entity tokens. The other verifies if the candidate entity tokens are contained within the mention tokens, a matching that we name as \texttt{sub-match}. We do this because we want to link to a broader entity, but never to a narrower entity in order to guarantee the validity of what is linked. Table \ref{table:exact_sub_match_examples} exemplifies some matching cases.

\begin{table}
\centering
\caption{\texttt{Exact-match} and \texttt{sub-match} examples.}
\begin{tabular}{|c|c|c|c|} 
\hline
\textbf{mention} & \textbf{entity} & \textbf{exact-match} & \textbf{sub-match}  \\ 
\hline
bright pink      & bright pink     & True                 & True                \\
bright pink      & pink            & False                & True                \\
pink             & bright pink     & False                & False               \\
bright pink      & light pink      & False                & False               \\
\hline
\end{tabular}
\label{table:exact_sub_match_examples}
\end{table}

The traditional ML models we explore are binary classifiers that predict a binary target for each mention-candidate pair. Since these models need numeric features as input, we build a set of features for each mention-candidate pair that are used by the binary classifier to make its prediction. These features can be split into two main types:
\begin{itemize}
    \item \textit{string similarity features}: as the name suggests, these features use string similarity measures. In particular, we use jaro-winkler\footnote{jaro-winkler - https://pypi.org/project/jaro-winkler/} to generate the similarity scores between the mention and the candidate entity tokens.
    \item \textit{semantic similarity features}: these features use vector representations (embeddings) for the mention and candidate entity, and then compute the cosine similarity between these representations. In order to generate specific embeddings for the fashion domain, we used a pre-trained BERT-based language model, fine-tuned on 236K textual descriptions from our product catalog, using Masked Language Modeling as in \cite{devlin2018bert}.
\end{itemize}

In order to keep the EL independent of the attribute types, not only do we generate the previous features using the mention and entity names, but we also use their types. This is in contrast with the rule-based models, in which we define a type mapper heuristic based on our domain expertise.


During inference, instead of predicting the binary label directly, we compute a linking score (i.e. the score associated with the positive target) and then rank all the candidate entities, for the same mention, by the linking score. Then, if the score of the top-rank candidate is larger than a specific threshold (that can be tuned during training), the attribute entity is linked.

\section{Experiments}
\label{section:experiments}
In this section, we outline the experiments conducted for the different components. In the first experiment, we fine-tune a BERT model in an NER task. In the second experiment, we train a few binary classifier models and compare them to our rule-based models, for the candidate entity disambiguation task. Both experiments use a small in-house dataset composed of textual product descriptions from the fashion domain.

\subsection{Named Entity Recognition}
As the world of fashion is a very complex domain with lots of categories and sub-categories arranged in a hierarchical manner, we focused only on \textit{Dresses} category for our experiments. We fine-tuned a pre-trained Transformer BERT-base-cased model~\cite{devlin2018bert} for NER using \textit{900} manually labeled product descriptions across 11 different tags, viz. \textit{shape}, \textit{hem}, \textit{color}, \textit{collar}, \textit{material}, \textit{fastening}, \textit{pockets}, \textit{neckline}, \textit{sleeve\_type}, \textit{pattern}, and \textit{detail}. The pre-trained model was trained on Wikipedia\footnote{https://huggingface.co/datasets/wikipedia} and the BookCorpus\footnote{https://huggingface.co/datasets/bookcorpus} datasets using Masked Language Modeling and Next Sentence Prediction tasks \cite{devlin2018bert}. There were 5702 and 1413 instances of named entities in the training and validation set, respectively, across different tags as shown in Table \ref{data:table_1}. 
Here, we used the Hugging Face library\footnote{Hugging Face library - www.huggingface.com} to fine-tune the pre-trained BERT model with the last linear layer on top configured to classify the named entities within our dataset. The model consisted of \emph{12} self-attention heads and \emph{12} hidden layers. Each hidden layer had \emph{768} units and there were \emph{512} maximum positional embeddings. The hidden layer units used \emph{gelu} activation units with a dropout probability of \emph{0.1}. We also used the BERT pre-trained tokenizer consisting of a vocabulary size of \emph{28996} tokens. Hugging Face enabled us to download the model with pre-trained weights. 

\begin{table}
\caption{Number of instances of named entities across the different entity types for both train and validation sets}
\resizebox{\columnwidth}{!}{%
\centering
\begin{tabular}{|l|c|c|c|c|c|c|c|c|c|c|c|c}
\hline
           & Detail & Color & Shape & Neckline & Pattern & Sleeve Type & Hem & Material & Fastening & Collar & Pockets  \\ 
\hline
Train   & 896    & 797   & 759   & 571      & 403     & 699          & 648 & 491      & 352       & 74     & 12       \\ 
\hline
Validation & 251    & 203   & 176   & 138      & 74      & 169          & 167 & 116      & 86        & 25     & 8        \\
\hline
\end{tabular}
}

\label{data:table_1}
\end{table}

We fine-tuned the above model on \emph{900} labeled samples, split into \emph{720} for training and \emph{180} for validation. Even though the dataset is small in size, we saw that having access to a pre-trained BERT-base-cased model provides us with the high accuracy necessary for the entity linking step. 

\subsection{Entity Linking}

The first component to experiment with was the \textit{candidate entity generation}. As explained in Section \ref{subsection:entity_linking}, we use full-text search over an inverted index of attribute values. In order to increase the recall set of the candidate attribute entities retrieved, we use fuzzy matching. We experimented with fuzziness scores in the range of $[0.5, 1.0]$, using increments of $0.1$, and manually tested a small list of attributes that should be retrieved for specific queries. We ended up choosing a \textit{fuzziness score} of $0.7$, which means that we match all terms having 70\% matching characters with the query.

Next, we trained and compared different models for the candidate entity disambiguation component. Since we needed labeled data to train (and test) the models, we manually annotated 32 instances, randomly sampled from the dataset used to train the NER model (see Table \ref{data:table_1}). Each instance contains a product textual description and the corresponding attribute mentions that should be recognized from the text. For each mention, we generate the candidate entities using the candidate entity generation module. We ended up with $2528$ samples,  where each sample is composed of a product textual description, an attribute mention, and an attribute candidate entity. Lastly, we manually labeled each sample using binary labels, with the positive samples representing the candidate attribute entities that should be linked, yielding $181$ ($7.2\%$) positive samples and $2347$ negative samples. The final dataset contains $132$ unique attribute mentions, covering the $11$ unique types described in Table \ref{data:table_1}, and $522$ unique attribute entities covering $33$ unique types of attributes (the top 10 attribute types with more samples are represented in Table  \ref{data:table_2}). We used a train/test split ratio of 9:1.

\begin{table}
\caption{Number of instances across the different attribute types}
\resizebox{\columnwidth}{!}{%
\centering
\begin{tabular}{|c|c|c|c|c|c|c|c|c|c|c|c}
\hline
    Material & Pattern & Neckline & Sleeve Type & Detail & Closure Type & Length & Color & Sleeve Length & Shape  \\ 
\hline
    444    & 397   & 280   & 248      & 179     & 174          & 137 & 134      & 91       & 70       \\ 
\hline
\end{tabular}
}
\label{data:table_2}
\end{table}

Regarding the candidate entity disambiguation models, we used the two rule-based models described in Section \ref{subsection:entity_linking} as baselines, and trained 3 different binary classifiers: a \texttt{logistic regression classifier}, a linear \texttt{support vector classifier} (SVC), and a \texttt{random forest classifier}. As metrics, we used the standard metrics for binary classification: \textit{precision}, \textit{recall} and \textit{f1-score}. We do not report the accuracy score due to the high unbalanced nature of the dataset.

\section{Results}
\label{section:results}

In this section, we report the accuracy numbers for the NER model and the results from the classification model for the \textit{candidate entity disambiguation} model. In addition, we also report a few enrichment statistics using the full system end to end.

\subsection{Named Entity Recognition}
To evaluate the performance of the fine-tuned BERT-base-cased model, we used the seqeval\footnote{seqeval - https://github.com/chakki-works/seqeval} framework in \emph{Python}. The fine-tuned BERT model achieved an accuracy of 89.75\% on the validation set. Due to the small size of the dataset, we had to stop the training of the model early enough to prevent overfitting.

\begin{figure}[htb]
  \centering
    \includegraphics[width=1.0\textwidth]{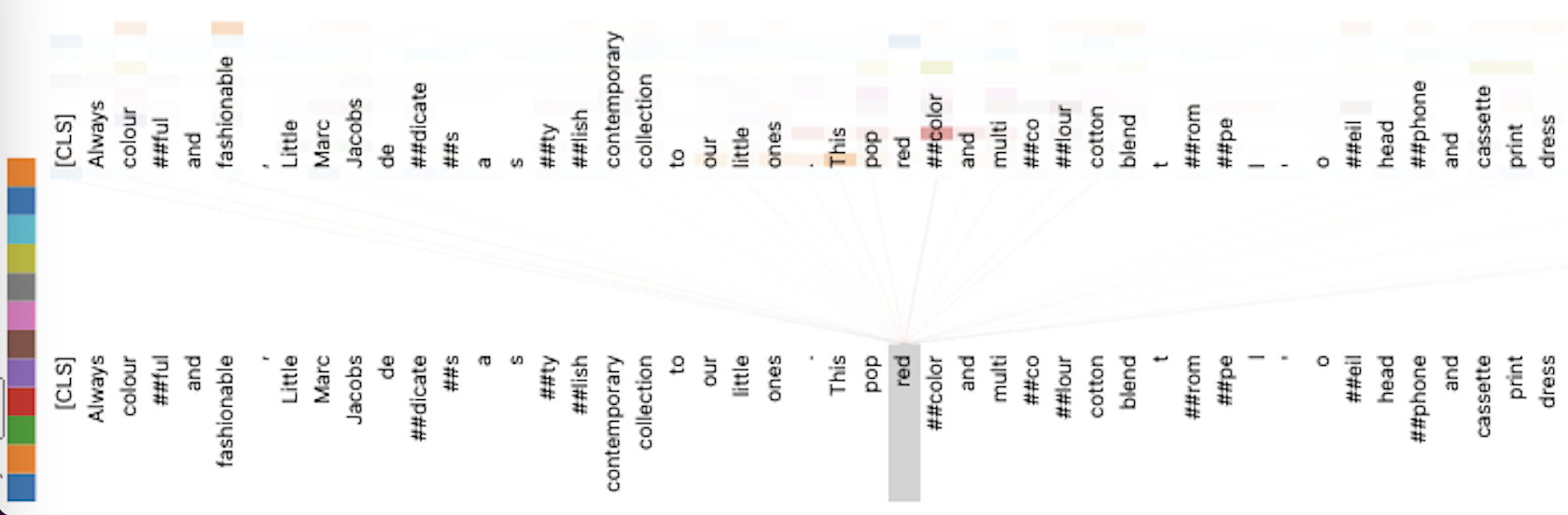}
      \caption{Visualization of the self-attention map for a sample product description.}
    \label{example:attention}
\end{figure}

We also visualized the weights of the self-attention layer showing the importance between the tokens in the input text. For example, as shown in Figure \ref{example:attention}, the token \textit{red} has a high correlation with tokens including \textit{fashionable} and \textit{color}. This way, we believe that the model learned an inner representation of the fashion domain language that can then be used to extract features useful for downstream tasks.

\subsection{Entity Linking}

We report the binary classification results on the test set, for the \textit{candidate entity disambiguation} task, in Table \ref{table:entity_disambiguation_results}. As expected, the \texttt{exact-match rule-based} model has a perfect \textit{precision}. Although there may be a few ambiguous cases in which there is an exact match with multiple attributes, we are constraining the linking based on the mention and entity types using a type mapper heuristic, and therefore disambiguating such cases. For example, consider the attribute mention \texttt{"long"}, labeled with the type \texttt{sleeve\_type} by NER (recall Table \ref{data:table_1}). Two exact-matched candidate entities are (\texttt{:long}, \texttt{:Length}) and (\texttt{:long}, \texttt{:Sleeve Length}). Due to the type mapper heuristic, only the (\texttt{:long}, \texttt{:Sleeve Length}) attribute entity is linked. The high \textit{precision} of the  \texttt{exact-match rule-based} model comes with the trade-off of a low \textit{recall} (only 56\% of the entities are linked).

\begin{table}
\centering
\caption{Binary classification results on the test set, for candidate entity disambiguation. The best results are highlighted in bold, with the second best in underline.}
\begin{tabular}{|l|c|c|c|} 
\hline
\textbf{Model}                 & \textbf{Precision} & {\textbf{Recall}} & \textbf{F1-Score}  \\ 
\hline
\texttt{exact-match}           & \textbf{1.00}       & 0.56                                & 0.72              \\
\texttt{sub-match}             & 0.78              & \uline{0.72}                        & 0.75               \\
\texttt{logistic-regression }           & \uline{0.94}      & 0.68                                & \uline{0.79}      \\
\texttt{svc}                            & 0.89              & 0.68                                & 0.77              \\
\texttt{random-forest} & 0.91              & \textbf{0.80}                        & \textbf{0.85}      \\
\hline
\end{tabular}
\label{table:entity_disambiguation_results}
\end{table}

In contrast, due to its less stringent matching mechanism, the \texttt{sub-match rule-based} model has a higher \textit{recall} ($0.72$), while maintaining a decent \textit{precision} ($0.783$), and thereby achieving an \textit{f1-score} of $0.75$.

Regarding the traditional ML binary classifiers, during training, we estimated the score thresholds for each model in order to keep a minimum \textit{precision} of $0.9$, using cross-validation experiments. The results reported in Table \ref{table:entity_disambiguation_results} correspond to the results on the test set, using the score threshold estimated during training. Although the \texttt{logistic regression} has the highest \textit{precision} ($0.94$), both the \texttt{support vector classifier} and the \texttt{random forest} models keep the minimum \textit{precision} requirement of $0.9$. The \texttt{random forest} classifier is the model that yields the best overall results, with the highest \textit{f1-score} (0.85), and therefore it is the model implemented in the candidate entity disambiguation component.

\subsection{End to End System}

The previous experiments and results describe the performance of both the NER and EL components individually. In order to estimate how the end to end system is performing, we applied the full system to a set of $650$ product textual descriptions, extracted from the \textit{dresses} category. On \textit{average}, for each product, we detect $7.79$ attribute mentions, we generate $12.8$ candidate entities for each mention, and end-up linking $5.68$ attribute entities. From all the entities linked, $52\%$ are exact matches.

In addition, there are an average of $2.63$ attribute mentions that are not linked per product. Table \ref{table:top_mentions_not_linked} reports the top 2 mentions without any links, that are detected in a large number of products. The \texttt{"straight hem"} mention represents a limitation of EL, which could easily be solved by improving the text-preprocessing, since we can find the attribute entity (\texttt{:straight hemline}, \texttt{:Hemline Style}) among the candidates. However, regarding the \texttt{"fitted waist"} mention, there are no relevant candidates. Since this mention occurs with high frequency in the products textual descriptions, it is a good indicator that we may need to create a corresponding new attribute entity in the FKG.

\begin{table}
\centering
\caption{Top 2 mentions without linked entities}
\begin{tabular}{l|c|c|c|} 
\textbf{Mention} & \texttt{"fitted waist"} & \texttt{"straight hem"} \\
\textbf{Mention Type}  & \texttt{shape}  & \texttt{hem}          \\
\textbf{Number of Products} & 63                    & 31                    \\

\end{tabular}

\label{table:top_mentions_not_linked}
\end{table}

Although the previous statistics demonstrate that we are enriching the products with an average of $5.68$ attributes, and the EL experimental results report a linking \textit{precision} of 0.9, we are still not completely certain about the correctness of the full enrichment, due to error propagation from the NER component to the downstream modules. Consider the following product textual description excerpt: \textit{"a bright blue dress, featuring a"}. A possible error would be for the NER to detect (\texttt{"blue"}, \texttt{color}), instead of (\texttt{"bright blue"}, \texttt{color}). This error would lead to the linking of the attribute (\texttt{:Blue}, \texttt{:Color}), instead of (\texttt{:Bright Blue}, \texttt{:Color}), in the EL component. Although not critical, such error propagation raises the need for improving the robustness of EL.

\section{Conclusion}
\label{section:conclusion}
In this paper, we have described how we applied existing methods to tackle the automatic enrichment of our KG within the fashion domain. We tackled the NER and the EL tasks and, despite the challenges brought by the lack of data and the specificity of the domain, we were able to augment our FKG, which is pivotal for the success of downstream tasks.
In particular, we have shown how we leveraged transfer learning approaches to fine-tune an NER model to the fashion domain using a small dataset. In addition, by using traditional ML models with simple features, we also built an EL component using a few labeled samples.

We are also confident that the embeddings obtained by fine-tuning the BERT-based model on the fashion domain capture the context and semantics specific to the domain. Hence, these domain-specific embeddings, which are a by-product of our enriching method, may enable several other downstream tasks.

However, our solution still presents several limitations. One of them is that the errors from the NER model may propagate into the EL step as a result of having a 2-step pipeline. This could be tackled by introducing further robustness matching steps. Having a jointly optimized process could also present advantages worth exploring. Nevertheless, such an approach requires the construction of a new dedicated dataset. So far, because of the efforts required for annotating data, our focus has been only on a subset of attributes for the \textit{Dresses} category, and we must scale this process to other attribute types and product categories in order to cover a larger set of our catalog. Therefore, we are focusing on defining the high-impact attributes for the business, and developing specific individual models targeting them, since it's crucial to guarantee high-quality results. However, for the remaining attributes, we may experiment with recent attribute value extraction methods, as in~\cite{scalingopentag2019}, that incorporate the attribute types directly as features and thereby are able to scale to thousands of attribute types, while requiring few annotated data points for each type. Furthermore, we feel there is still scope for improvement on entity disambiguation, across multiple levels, namely on text pre-processing, better feature engineering by generating richer features, and increasing model complexity.

Besides the above limitations, and as an additional direction for future work, we feel we can take advantage of the NIL clustering to detect new attributes in order to automatically update the ever-growing and ever-changing taxonomy of FKG.

\subsubsection*{Acknowledgements}
  The authors would like to thank all team members of the FKS team at Farfetch. Their support was crucial for the conducted research.

\bibliographystyle{plain}
\bibliography{references}

\begin{thebibliography}{10}

\bibitem{NELKGsurvey2020}
T.~Al-Moslmi, M.~O. Gallofre, A.~L. Opdahl, and C.~Veres.
\newblock Named entity extraction for knowledge graphs: A literature overview.
\newblock {\em IEEE Access}, 8:32862--32881, 2020.

\bibitem{cao3}
N.~De Cao, W.~Aziz, and I.~Titov.
\newblock Highly parallel autoregressive entity linking with discriminative
  correction.
\newblock {\em CoRR}, abs/2109.03792, 2021.

\bibitem{cao}
N.~De Cao, G.~Izacard, S.~Riedel, and F.~Petroni.
\newblock Autoregressive entity retrieval.
\newblock {\em CoRR}, abs/2010.00904, 2020.

\bibitem{cao2}
N.~De Cao, L.~Wu, K.~Popat, M.~Artetxe, N.~Goyal, M.~Plekhanov, L.~Zettlemoyer,
  N.~Cancedda, S.~Riedel, and F.~Petroni.
\newblock Multilingual autoregressive entity linking.
\newblock {\em CoRR}, abs/2103.12528, 2021.

\bibitem{ycao2018}
Y.~Cao, L.~Hou, J.~Li, and Z.~Liu.
\newblock Neural collective entity linking.
\newblock In {\em Proc. 27th Int. Conf. Comput. Linguistics}, page 675–686,
  2018.

\bibitem{conneau-etal-2020-roberta}
Alexis Conneau, Kartikay Khandelwal, Naman Goyal, Vishrav Chaudhary, Guillaume
  Wenzek, Francisco Guzm{\'a}n, Edouard Grave, Myle Ott, Luke Zettlemoyer, and
  Veselin Stoyanov.
\newblock Unsupervised cross-lingual representation learning at scale.
\newblock In {\em Proceedings of the 58th Annual Meeting of the Association for
  Computational Linguistics}, pages 8440--8451, July 2020.

\bibitem{devlin2018bert}
J.~Devlin, M.W. Chang, K.~Lee, and K.~Toutanova.
\newblock Bert: Pre-training of deep bidirectional transformers for language
  understanding.
\newblock {\em arXiv preprint arXiv:1810.04805}, 2018.

\bibitem{hoffart}
J.~Hoffart, M.~A. Yosef, I.~Bordino, H.~F{\"u}rstenau, M.~Pinkal, M.~Spaniol,
  B.~Taneva, S.~Thater, and G.~Weikum.
\newblock Robust disambiguation of named entities in text.
\newblock In {\em Proceedings of the 2011 Conference on Empirical Methods in
  Natural Language Processing}, pages 782--792, Edinburgh, Scotland, UK., July
  2011. Association for Computational Linguistics.

\bibitem{biLSTM-CRF2015}
Z.~Huang, W.~Xu, and K.~Yu.
\newblock Bidirectional lstm-crf models for sequence tagging.
\newblock {\em CoRR}, abs/1508.01991, 2015.

\bibitem{kolitsas2018}
N.~Kolitsas, O.~E. Ganea, and T.~Hofmann.
\newblock End-to-end neural entity linking.
\newblock In {\em Proc. 22nd Conf. Comput. Natural Lang. Learn.}, page
  519–529, 2018.

\bibitem{lample-etal-2016-neural}
G.~Lample, M.~Ballesteros, S.~Subramanian, K.~Kawakami, and C.~Dyer.
\newblock Neural architectures for named entity recognition.
\newblock In {\em Proceedings of the 2016 Conference of the North {A}merican
  Chapter of the Association for Computational Linguistics: Human Language
  Technologies}, pages 260--270, 2016.

\bibitem{letitov2018}
P.~Le and I.~Titov.
\newblock Improving entity linking by modeling latent relations between
  mentions.
\newblock In {\em Proc. 56th Annu. Meeting Assoc. Comput. Linguistics}, page
  1595–1604, 2018.

\bibitem{marcelino2018hierarchical}
Jos{\'e} Marcelino, Jo{\~a}o Faria, Lu{\'\i}s Ba{\'\i}a, and Ricardo~Gamelas
  Sousa.
\newblock A hierarchical deep learning natural language parser for fashion.
\newblock {\em arXiv preprint arXiv:1806.09511}, 2018.

\bibitem{martins2019}
P.~H. Martins, Z.~Marinho, and A.~F.~T. Martins.
\newblock Joint learning of named entity recognition and entity linking.
\newblock In {\em Proc. 57th Annu. Meeting Assoc. Comput. Linguistics, Student
  Res. Workshop}, page 190–196, 2019.

\bibitem{mulang}
I.~O. Mulang, K.~Singh, C.~Prabhu, A.~Nadgeri, J.~Hoffart, and J.~Lehmann.
\newblock Evaluating the impact of knowledge graph context on entity
  disambiguation models.
\newblock {\em CoRR}, abs/2008.05190, 2020.

\bibitem{ravi}
M.~Ravi, K.~Singh, I.~O. Mulang, S.~Shekarpour, J.~Hoffart, and J.~Lehmann.
\newblock {CHOLAN:} {A} modular approach for neural entity linking on wikipedia
  and wikidata.
\newblock {\em CoRR}, abs/2101.09969, 2021.

\bibitem{bm25}
S.~E. Robertson and S.~Walker.
\newblock Some simple effective approximations to the 2-poisson model for
  probabilistic weighted retrieval.
\newblock In {\em Proceedings of the 17th Annual International ACM SIGIR
  Conference on Research and Development in Information Retrieval}, SIGIR '94,
  page 232–241, Berlin, Heidelberg, 1994. Springer-Verlag.

\bibitem{NELDLsurvey1}
{\"{O}}.~Sevgili, A.~Shelmanov, M.~Y. Arkhipov, A.~Panchenko, and C.~Biemann.
\newblock Neural entity linking: {A} survey of models based on deep learning.
\newblock {\em CoRR}, abs/2006.00575, 2020.

\bibitem{NELDLsurvey2}
W.~Shen, Y.~Li, Y.~Liu, J.~Han, J.~Wang, and X.~Yuan.
\newblock Entity linking meets deep learning: Techniques and solutions.
\newblock {\em CoRR}, abs/2109.12520, 2021.

\bibitem{ShenSurvey2015}
W.~Shen, J.~Wang, and J.~Han.
\newblock Entity linking with a knowledge base: Issues, techniques, and
  solutions.
\newblock {\em IEEE Transactions on Knowledge and Data Engineering},
  27(2):443--460, 2015.

\bibitem{Transformer2017}
A.~Vaswani, N.~Shazeer, N.~Parmar, J.~Uszkoreit, L.~Jones, A.~Gomez, L.~Kaiser,
  and I.~Polosukhin.
\newblock Attention is all you need.
\newblock In {\em Advances in Neural Information Processing Systems},
  volume~30, 2017.

\bibitem{scalingopentag2019}
H.~Xu, W.~Wang, X.~Mao, X.~Jiang, and M.~Lan.
\newblock Scaling up open tagging from tens to thousands: Comprehension
  empowered attribute value extraction from product title.
\newblock In {\em Proceedings of the 57th Conference of the Association for
  Computational Linguistics, {ACL} 2019}, pages 5214--5223, 2019.

\bibitem{NERsurvey2019}
V.~Yadav and S.~Bethard.
\newblock A survey on recent advances in named entity recognition from deep
  learning models.
\newblock {\em CoRR}, abs/1910.11470, 2019.

\bibitem{yamada}
I.~Yamada and H.~Shindo.
\newblock Pre-training of deep contextualized embeddings of words and entities
  for named entity disambiguation.
\newblock {\em CoRR}, abs/1909.00426, 2019.

\bibitem{opentag2018}
G.~Zheng, S.~Mukherjee, X.~Luna Dong, and F.~Li.
\newblock Opentag: Open attribute value extraction from product profiles.
\newblock In {\em Proceedings of the 24th ACM SIGKDD International Conference
  on Knowledge Discovery \& Data Mining}, KDD '18, page 1049–1058, 2018.

\end{thebibliography}

\end{document}